\documentclass[aps,nofootinbib,superscriptaddress,twocolumn,tightenlines]{revtex4}
\usepackage{amsmath,amssymb,graphicx,epsfig,hyperref,breakurl,color,bm,subfigure,slashed,lipsum,multirow}
\allowdisplaybreaks[4]

\AtBeginEnvironment{align}{\small}
\AtBeginEnvironment{eqnarray}{\small}
\begin{document}
\newcommand{\psl}{ p \hspace{-1.8truemm}/ }
\newcommand{\nsl}{ n \hspace{-2.2truemm}/ }
\newcommand{\vsl}{ v \hspace{-2.2truemm}/ }
\newcommand{\epsl}{\epsilon \hspace{-1.8truemm}/\,  }
\title{ Large CP violation in $\Lambda_b\rightarrow \Lambda D$ decays and extraction of the Cabibbo-Kobayashi-Maskawa angle  $\gamma$ }

\author{Zhou Rui}
\email{jindui1127@126.com}
\affiliation{Department of Physics, Yantai University, Yantai 264005, China}

\author{Zhi-Tian Zou}
\email{zouzt@ytu.edu.cn}
\affiliation{Department of Physics, Yantai University, Yantai 264005, China}

\author{Ya Li}
\email{liyakelly@163.com}
\affiliation{Department of Physics and Institute of Theoretical Physics, Nanjing Normal University,Nanjing 210023, China}

\author{Ying Li}
\email{liying@ytu.edu.cn}
\affiliation{Department of Physics, Yantai University, Yantai 264005, China}
\date{\today}
\begin{abstract}

Motivated by the first observation of CP violation in $b$-baryon decays, the search for baryonic decays exhibiting large CP violation will be a primary focus in the coming years. We propose that significant CP-violating effects exist in the decay $\Lambda_b \to \Lambda D$, where $D$ denotes a CP eigenstate of the $D^0 - \bar{D}^0$ system. The predicted CP asymmetries for both the CP-even and CP-odd modes can reach magnitudes as large as $50\%$, making these decays promising targets for measurement at the LHCb experiment. Additionally, we predict for the first time several nonzero CP-violating observables associated with angular distribution parameters, providing valuable complementary information in the search for CP violation in baryon decays. Furthermore, we propose a novel strategy to extract the CKM angle $\gamma$ by combining data on angular distribution parameters and decay rates from the relevant channels. We emphasize that $\Lambda_b \rightarrow \Lambda D$ decays are among the most promising candidates for determining $\gamma$ in the baryon sector. Our findings may offer new insights for future theoretical and experimental investigations.


\end{abstract}

\pacs{13.25.Hw, 12.38.Bx, 14.40.Nd }

\maketitle

{\it Introduction.---}
In 2025, the LHCb Collaboration reported the first observation of baryonic CP violation (CPV) in the decay \(\Lambda_b \rightarrow p K^- \pi^+ \pi^-\), achieving a significance of 5.2 standard deviations~\cite{LHCb:2025ray}. 
Previously, evidence for CPV had been observed in the three-body decay 
$\Lambda_b \rightarrow \Lambda K^+K^-$~\cite{LHCb:2024yzj}, and ongoing studies of CPV have been conducted in various three- and four-body decay modes of $\Lambda_b$ and $\Xi_b$~\cite{LHCb:2014yin, LHCb:2014nhe, LHCb:2018fpt, LHCb:2019oke, LHCb:2019jyj,LHCb:2025ozp}. However, measurements of direct CPV in $b$-baryon decays have generally yielded  values below $10\%$, suggesting that CPV originating from tree-penguin interference is difficult to be significant within the Standard Model (SM). This is due to the suppressed penguin-to-tree amplitude ratio and cancellations among  partial wave CPVs~\cite{Han:2024kgz}. Consequently, the search for baryonic processes exhibiting large CPV is expected to be a primary focus in the coming years~\cite{Shen:2023eln, He:2025msg}.

In this Letter, we propose a potential decay channel with sizable CPV: \(\Lambda_b \rightarrow \Lambda D\), where \(D\) denotes an admixture of the \(D^0\) and \(\bar{D}^0\) states. We demonstrate that large CPV exists when \(D\) is a CP eigenstate, defined as \(D_\pm = \frac{1}{\sqrt{2}}(D^0 \pm \bar{D}^0)\). 
This CPV originates from tree-tree interference between the amplitudes \(\mathcal{M}(\Lambda_b \rightarrow \Lambda D^0)\) and \(\mathcal{M}(\Lambda_b \rightarrow \Lambda \bar{D}^0)\). 
The two interfering amplitudes have comparable magnitudes of \(\mathcal{O}(\lambda^3)\), where \(\lambda \approx 0.23\) is the sine of the Cabibbo angle~\cite{Wolfenstein:1983yz}. Such an interference mechanism is of the tree-tree type, differing from the conventional tree-penguin interference observed in decays such as \(\Lambda_b \rightarrow p \pi\)~\cite{Han:2024kgz}. As pointed out in Ref.~\cite{Han:2024kgz}, the cancellation among partial-wave CP asymmetries is traced to a relative sign between the \(S\)- and \(P\)-wave penguin contributions arising from the \((S+P)(S-P)\) operator structure. However, for tree-tree interference, where all contributing tree operators share the same \((V-A)(V-A)\) structure, such cancellation 
is naturally absent. We will later uncover that the partial-wave CP asymmetries in the decays under consideration share the same sign, leading to a coherent enhancement of the total CPV. 

To observe CPV, the decay amplitude must receive contributions from at least two distinct pathways that differ in both strong and weak phases. The weak and strong phase differences, together with the ratio between the two interfering amplitudes, are key factors in determining the magnitude of CPV. In the decays considered here, the weak phase difference is simply the Cabibbo–Kobayashi–Maskawa (CKM) angle $\gamma$. A precise evaluation of the strong phases requires comprehensive QCD calculations of various nonfactorizable topological diagrams, such as internal $W$-emission and $W$-exchange diagrams~\cite{Cheng:1996cs}, which provide significant strong phases. 
The ratio of the relevant CKM matrix elements between the two amplitudes is $|V_{ub}V_{cs}/V_{cb}V_{us}|\sim0.37$, which alone might not suggest very large CPV. For example, in the corresponding mesonic decays $B^-\to K^- D_{\pm}$, the amplitude ratio is further suppressed to about $0.1$ due to a color suppression factor~\cite{Belle:2006cuz}, and the measured CP asymmetries are of the order of $10\%$~\cite{ParticleDataGroup:2024cfk}. However, the situation in $\Lambda_b\to \Lambda D_{\pm}$ decays is distinct due to their more complex dynamics. As we will show later through a full QCD calculation, the dynamical enhancement from internal $W$-emission diagrams in $\mathcal{M}(\Lambda_b \to \Lambda \bar{D}^0)$ increases the amplitude ratio to approximately unity. These combined effects lead to large CPV in the $\Lambda_b \to \Lambda D_\pm$ decays.

The large CPVs in the \(\Lambda_b \to \Lambda D_\pm\) decays suggest that these channels could serve as ``golden modes" for extracting the CKM angle \(\gamma\) in the baryon sector. 
Currently, direct \(\gamma\) measurements come mainly from \(B\to D\) decays via \(D^0\)-\(\bar D^0\) mixing~\cite{Belle:2024knt, BaBar:2013caj, LHCb:2024yxi,LHCb:2025fom}, while broader strategies involving \(D^*\) and \(D_s\) final states have been studied by the Heavy Flavor Averaging Group (HFLAV)~\cite{FlavourLatticeAveragingGroupFLAG:2024oxs}. 
Direct \(\gamma\) measurements suffer from larger uncertainties than indirect CKM fits~\cite{Charles:2015gya, UTfit:2022hsi}, so improving their precision is essential to better constrain BSM contributions to CPV. Baryon modes such as \(\Lambda_b \to \Lambda D_\pm\) provide complementary constraints, and comparing \(\gamma\) values extracted from different \(b\)-hadron species. 
Although earlier works~\cite{Dunietz:1992ti, Giri:2001ju, Zhang:2021sit, Fayyazuddin:1998aa, Geng:2022osc} have explored baryonic extractions, the present analysis offers a complete QCD calculation of the strong phase, which will be valuable for future experimental determinations of \(\gamma\).

\begin{figure}[!htbh]
	\begin{center}
	 \centerline{\epsfxsize=5.8cm \epsffile{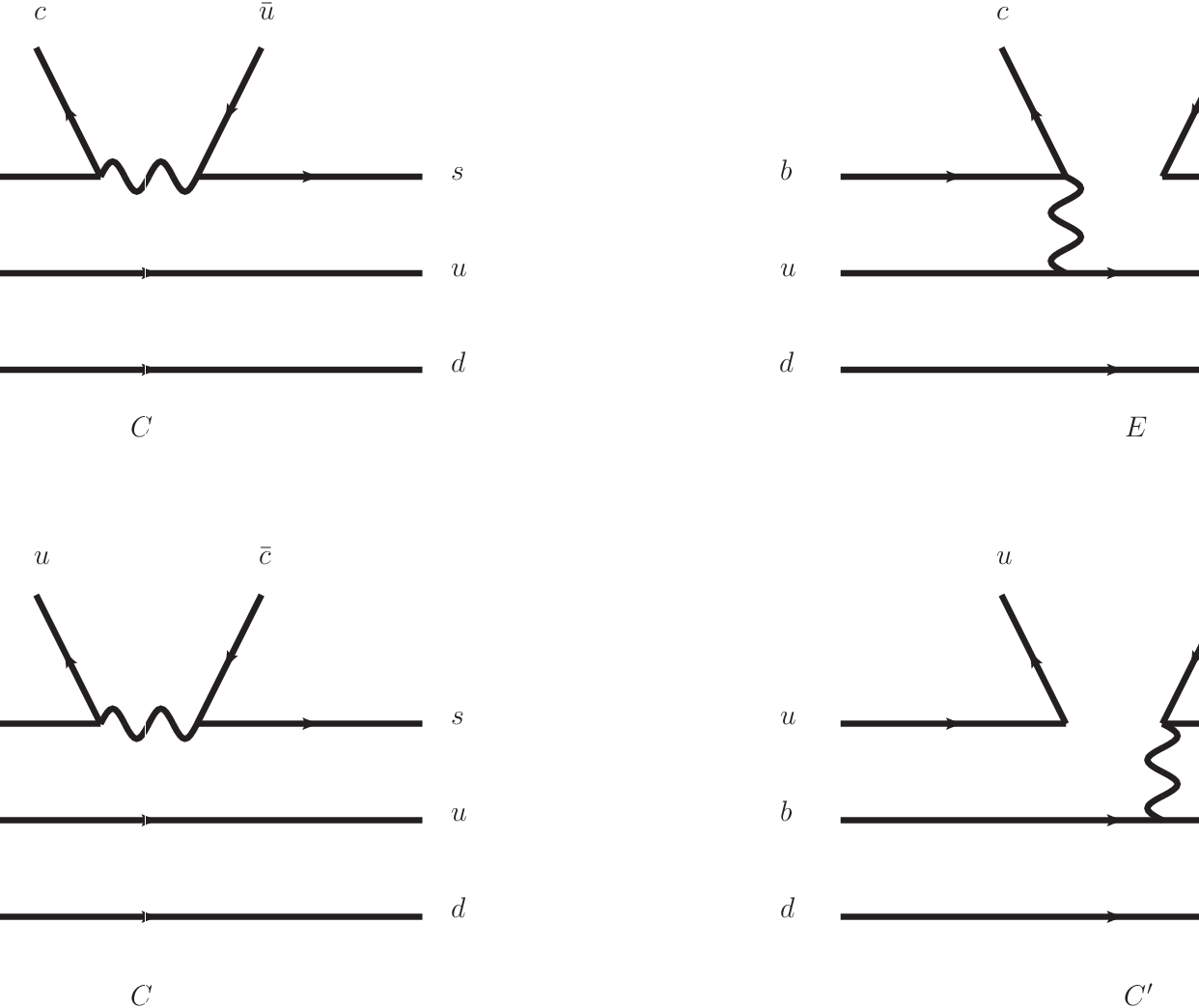}}
\caption{Decay topologies referred to as color-suppressed ($C$),  $W$-exchange ($E$), and internal $W$-emission diagrams $C'$.} 
		\label{fig:feym}
	\end{center}
\end{figure}

{\it Decay amplitudes and $CP$ asymmetries in $\Lambda_b\to \Lambda D$.---}
Figure~\ref{fig:feym} displays the topological diagrams for the $\Lambda_b \to \Lambda D^0(\bar{D}^0)$ decays, where $C$, $E$, and $C'$ denote the color-suppressed $W$-emission, $W$-exchange, and internal $W$-emission diagrams, respectively. The $C$ diagram contributes to both the $D^0$ and $\bar{D}^0$ final states, while $E$ ($C'$) contributes only to the former (latter). According to the power counting rules in soft-collinear effective theory (SCET)~\cite{Mantry:2003uz,Leibovich:2003tw}, $C$, $C'$, and $E$ are all suppressed relative to the external $W$-emission diagram $T$ by $\Lambda_{\mathrm{QCD}}/m_b$, where $\Lambda_{\mathrm{QCD}}$ is the QCD scale and $m_b$ the $b$-quark mass; hence, their contributions are expected to be comparable at leading order in the $\Lambda_{\mathrm{QCD}}/m_b$ expansion. We will show that including higher-twist corrections for $\Lambda_b \to \Lambda D$ leads to the amplitude hierarchy $C \sim C' \gg E$. The constructive interference between $C$ and $C'$ significantly enhances the $\Lambda_b \to \Lambda \bar{D}^0$ amplitude, yielding amplitude ratios of order unity for both $S$- and $P$-wave components. This is the main origin of the sizable CP violation observed in $\Lambda_b \to \Lambda D$ decays. 

Angular momentum conservation allows both $S$-wave and $P$-wave amplitudes to contribute to the $\Lambda_b \to \Lambda D$ decay. The corresponding Lorentz-invariant amplitude  can be written as
\begin{eqnarray}\label{eq:MM}
\mathcal{M}(\Lambda_b\rightarrow \Lambda D)= i \bar{u}_{\Lambda}(A+B\gamma_5)u_{\Lambda_b},
\end{eqnarray}
where $u_{\Lambda_b(\Lambda)}$ is the Dirac spinor for $\Lambda_b(\Lambda)$ baryon. The $S$- and $P$-wave amplitudes are given by
\begin{eqnarray}\label{eq:SP}
S=\sqrt{Q_+}A,\quad P=\sqrt{Q_-}B,
\end{eqnarray}
where $Q_{\pm}=(M\pm m)^2-m_D^2$ with $M(m)$ being the mass of the initial (final) state baryon and $m_D$ the $D$ meson mass. The decay width is expressed as
\begin{eqnarray}\label{eq:width}
\Gamma= \frac{p_c}{8\pi M^2}\mathcal{A},
\end{eqnarray}
with $\mathcal{A}=|S|^2+|P|^2$. Here, $p_c$ is the momentum of $\Lambda$ in the rest frame of $\Lambda_b$ baryon. 
For the convenience of the subsequent discussion, we write the two partial-wave amplitudes as their respective magnitude and phase:
\begin{eqnarray}\label{eq:SP1}
S_D=|S_D|e^{i \delta_{S_D}},\quad P_D=|P_D|e^{i \delta_{P_D}},
\end{eqnarray}
for the $D^0$ final state, and
\begin{eqnarray}\label{eq:SP2}
S_{\bar{D}}=|S_{\bar{D}}|e^{i(\delta_{S_{\bar{D}}}-\gamma)},\quad P_{\bar{D}}=|P_{\bar{D}}|e^{i( \delta_{P_{\bar{D}}}-\gamma)},
\end{eqnarray}
for the $\bar{D}^0$ final state. 

Now consider the CP eigenstates, $D_\pm$. The corresponding partial-wave amplitudes can be directly written as
\begin{eqnarray}\label{eq:SPcp}
S_{\pm}=\frac{1}{\sqrt{2}} (S_{D}\pm S_{\bar{D}}),\quad P_{\pm}=\frac{1}{\sqrt{2}} (P_{D}\pm P_{\bar{D}}).
\end{eqnarray}
The amplitudes for the  $CP$-conjugate processes are given by
\begin{eqnarray}\label{eq:SPcpb}
\bar{S}_{\pm}=\frac{1}{\sqrt{2}} (\bar{S}_{D}\pm \bar{S}_{\bar{D}}),\quad \bar{P}_{\pm}=\frac{1}{\sqrt{2}} (\bar{P}_{D}\pm \bar{P}_{\bar{D}}),
\end{eqnarray}
where the quantities with a bar denote the counterparts of charge-conjugated processes. Their expressions can be obtained from Eqs.~(\ref{eq:SP1}) and~(\ref{eq:SP2}) by flipping the weak phase ($\gamma\rightarrow -\gamma$), while leaving the strong phases unchanged. 
Using these partial-wave amplitudes, we can define the partial-wave CP  asymmetries, $A_{\pm}^{S(P)}$, as~\cite{Han:2025tvc}
{\small
\begin{align}
\label{eq:CPV1}
A_{\pm}^{S}&\equiv\frac{|S_\pm|^2-|\bar{S}_\pm|^2}{|S_\pm|^2+|\bar{S}_\pm|^2}=\frac{\pm2r_S\sin{\gamma}\sin\delta_{S_DS_{\bar D}}}{1+r_S^2\pm2r_S\cos{\gamma}\cos\delta_{S_DS_{\bar D}}},\nonumber\\ 
A_{\pm}^{P}&\equiv\frac{|P_\pm|^2-|\bar{P}_\pm|^2}{|P_\pm|^2+|\bar{P}_\pm|^2}=\frac{\pm2r_P\sin{\gamma}\sin\delta_{P_DP_{\bar D}}}{1+r_P^2\pm 2r_P\cos{\gamma}\cos\delta_{P_DP_{\bar D}}},  
\end{align}}
where the magnitudes of the partial-wave amplitude ratio are $r_{S(P)}=\left|\frac{S(P)_{\bar{D}}}{S(P)_D}\right|$ and strong phase differences are $\delta_{S_DS_{\bar{D}}(P_DP_{\bar{D}})}=\delta_{S_{\bar{D}}(P_{\bar{D}})}-\delta_{S_D(P_D)}$.  Similarly,  we have the global $CP$ asymmetry,
{\small
\begin{align}\label{eq:CPV}
A_{CP}(\Lambda_b\rightarrow \Lambda D_{\pm})&\equiv\frac{|S_{\pm}|^2+|P_{\pm}|^2-|\bar{S}_{\pm}|^2-|\bar{P}_{\pm}|^2}{|S_{\pm}|^2+|P_{\pm}|^2+|\bar{S}_{\pm}|^2+|\bar{P}_{\pm}|^2},
\end{align}}
which can be written as a weighted combination of the partial-wave CP asymmetries
\begin{eqnarray}\label{eq:CPV2}
A_{CP}(\Lambda_b\rightarrow \Lambda D_{\pm})=\kappa^{\pm}A_{\pm}^S+(1-\kappa^{\pm})A_{\pm}^P,
\end{eqnarray}
with the weighted coefficients $\kappa^{\pm}=\frac{R^{\pm}_{S}}{R^{\pm}_S+R^{\pm}_P}$, and
\begin{eqnarray}
R^{\pm}_S&=&|S_D|^2[1+r_S^2\pm 2r_S\cos{\gamma}\cos\delta_{S_DS_{\bar D}}], \\
R^{\pm}_P&=&|P_D|^2[1+r_P^2\pm 2r_P\cos{\gamma}\cos\delta_{P_DP_{\bar D}}].
\end{eqnarray}
From Eq.~\ref{eq:CPV2}, one can see that if the different partial-wave CPVs share the same sign, the overall direct CP asymmetry  increases.

Accurate estimation of CPVs requires reliable partial-wave amplitudes including strong phases. Naive factorization~\cite{Wirbel:1985ji} is inapplicable to color-suppressed decays~\cite{Neubert:2001sj} and yields real amplitudes without relative phases, hence no CPV. Therefore, nonfactorizable diagrams (Fig.~\ref{fig:feym}) are needed to provide the necessary strong phases. In this work, we evaluate the topological diagrams $C'$ and $E$ for the first time using the perturbative QCD (PQCD) formalism based on $k_T$-factorization~\cite{Li:1994cka}, which avoids endpoint singularities. 
Further computational details are given in the Supplemental Material~\cite{SM}.

\begin{table}[!htbh]
\caption{Contributions from the various topologies to the partial-wave amplitudes ($10^{-6}$ GeV) without the CKM matrix elements for the $\Lambda_b\rightarrow \Lambda D^0$ decay. 
The strong phases are given in rad.}
	\label{tab:amplitude1}
	\begin{tabular}[t]{lcccccc}
	\hline\hline
 Topology & $S_D$ & $|S_D|$ & $\delta_{S_D}$ & $P_D$ & $|P_D|$ & $\delta_{P_{D}}$\\ \hline
  $C_f$              & 1.8       &1.8  &0               & 1.8 &1.8  &0\\
  $C_{nf}$           & 3.0-4.3\textit{i}  &5.2  &-0.96           & 1.8-2.9\textit{i} &3.4   &-1.02\\
  $C=C_f+C_{nf}$     & 4.8-4.3\textit{i}  &6.4  &-0.73           & 3.6-2.9\textit{i} &4.6  &-0.68\\
  $E  $              &-0.3-0.02\textit{i} &0.3  &-3.07           & -0.3+0.5\textit{i}&0.6   &2.11\\
  Total              &4.5-4.3\textit{i}   &6.2  &-0.76           &3.3-2.4\textit{i}  &4.1   &-0.63\\
		\hline\hline
	\end{tabular}
\end{table}

\begin{table}[!htbh]
\caption{Same as Table~\ref{tab:amplitude1} but for the $\Lambda_b\rightarrow \Lambda \bar{D}^0$ decay. }	\label{tab:amplitude2}
\begin{tabular}[t]{lcccccc}
\hline\hline
 Topology & $S_{\bar D}$ & $|S_{\bar D}|$ & $\delta_{S_{\bar D}}$ & $P_{\bar D}$ & $|P_{\bar D}|$ & $\delta_{P_{\bar D}}$\\ \hline
  $C_f$              & 1.8       &1.8  &0               & 1.8        &1.8    &0\\
  $C_{nf}$           & 0.3-6.4\textit{i}  &6.4  &-1.52           & -0.4-5.0\textit{i}  &5.0    &-1.65\\
  $C=C_f+C_{nf}$     & 2.1-6.4\textit{i}  &6.7 &-1.25            & 1.4-5.0\textit{i}   &5.2    &-1.30\\
  $C'  $             &1.0-5.9\textit{i}   &6.0  &-1.40           & 0.8-6.1\textit{i}   &6.2   &-1.44\\
  Total              &3.1-12.3\textit{i}  &12.7 &-1.32           & 2.2-11.1\textit{i}  &11.3   &-1.38\\
		\hline\hline
	\end{tabular}
\end{table}

Tables~\ref{tab:amplitude1} and~\ref{tab:amplitude2} list the partial-wave amplitudes (excluding CKM factors) for $\Lambda_b \to \Lambda D^0$ and $\Lambda_b \to \Lambda \bar{D}^0$ (central values only), with total amplitudes in the last row. 
The gluon exchanges between four-quark operators and spectator quarks (Fig.~\ref{fig:feym}) subdivides the $C$ topology into factorizable $C_f$ and nonfactorizable $C_{nf}$, distinguished by whether the meson participates in the gluon exchange. Independent PQCD evaluation of each diagram ensures unambiguous separation.  
The decays are clearly dominated by nonfactorizable contributions, indicating the inadequacy of factorization. 
The similar phases of $C$ and $C'$ lead to constructive interference, making the amplitude for $\Lambda_b \to \Lambda \bar{D}^0$ of the same order as that for $\Lambda_b \to \Lambda D^0$. Including the appropriate CKM factors yields
\begin{eqnarray}\label{eq:ratiosp}
 r_S&=&\left|\frac{V_{ub}V^*_{cs}}{V_{cb}V^*_{us}}\right|\times\left|\frac{S_{\bar D}}{S_D}\right|=0.76, \nonumber\\
 r_P&=&\left|\frac{V_{ub}V^*_{cs}}{V_{cb}V^*_{us}}\right|\times\left|\frac{P_{\bar D}}{P_D}\right|=1.05. 
\end{eqnarray}
The ratios between the interfering $\bar{D}^0$ and $D^0$ amplitudes are close to 1 for both $S$- and $P$-wave components.

\begin{table}[!htbh]
	\caption{ Comparison of theoretical predictions on the branching fractions $(10^{-6})$ without the nonfactorizable contribution.} 
	\label{tab:obser1}
	\begin{tabular}[t]{lcccc}
	\hline\hline
    Mode  & This work   & ~\cite{Zhu:2018jet} &\cite{Giri:2001ju} &\cite{Geng:2022osc} \\ \hline
   $D^0$        &$3.64^{+2.33}_{-1.53}$  &$3.37^{+0.33+0.42+0.67}_{-0.19-0.47-0.23}$   &$4.56$   &$6.6\pm0.6$\\
   $\bar{D}^0$  &$0.49^{+0.31}_{-0.21}$  &$0.478^{+0.060+0.103+0.061}_{-0.027-0.108-0.047}$    &$0.829$  &$0.9\pm0.1$\\
		\hline\hline
	\end{tabular}
\end{table}

\begin{table}[!htbh]
	\caption{The predicted branching fractions, direct $CP$ asymmetries, and the weighted partial-wave $CP$ asymmetries of $\Lambda_b\rightarrow \Lambda D$ decays.}
	\label{tab:obser}
	\begin{tabular}[t]{lcccc}
	\hline\hline
   $D$  & $\mathcal{B}(10^{-5})$ & $A_{CP}$   & $\kappa A^S_{CP}$ &$(1-\kappa) A^P_{CP}$ \\ \hline
 $D^0$        &$3.1^{+1.8}_{-0.8}$  &0   &0   &0\\
 $\bar{D}^0$  &$2.3^{+1.1}_{-0.7}$  &0   &0   &0\\
 $D_{+}$      &$1.9^{+1.1}_{-0.4}$  &$-0.44^{+0.10}_{-0.03}$   &$-0.25^{+0.08}_{-0.01}$   &$-0.19^{+0.04}_{-0.04}$\\
 $D_{-}$      &$3.5^{+1.9}_{-1.0}$  &$0.72^{+0.04}_{-0.14}$
  &$0.42^{+0.01}_{-0.11}$     &$0.30^{+0.05}_{-0.06}$ \\
		\hline\hline
	\end{tabular}
\end{table}

To examine the factorization ansatz, we summarize the branching fractions for the $\Lambda_b \to \Lambda D$ decays, excluding the nonfactorizable contribution, in Table~\ref{tab:obser1}. For comparison, we also present the currently available predictions based on naive factorization. The quoted theoretical uncertainties in our calculations arise from the hadronic parameters in the LCDAs as well as the hard scale. Our results are consistent with the estimations from the diquark approximation~\cite{Zhu:2018jet}.  
Without nonfactorizable contributions, no relative strong phase exists, and hence the factorization ansatz produces no CPV.  

Including nonfactorizable contributions, Table~\ref{tab:obser} gives the predicted branching fractions, direct CP asymmetries, and weighted partial-wave CP asymmetries for $\Lambda_b \to \Lambda D$. As expected, the branching fractions are enhanced by an order of magnitude to $\sim 10^{-5}$, comparable to those observed in analogous color-suppressed $B$ decays, e.g., $\bar{B}^0 \to D^0\bar{K}^0$~\cite{BaBar:2006ftr} and $\bar{B}^0_s \to D^0\phi$~\cite{LHCb:2023fqx}.

Even in the presence of nonzero strong phases, the CPVs of the $D^0$ and $\bar{D}^0$ channels  remain zero due to the vanishing weak phase difference. However, for both CP-even and CP-odd eigenstates, we observe sizeable CPVs with similar magnitudes, as shown in Table~\ref{tab:obser}. The opposite signs of these asymmetries can be easily understood from Eq.~(\ref{eq:CPV1}). This pattern is analogous to the measured CPVs in  $B\rightarrow K D_{\pm}$ decays~\cite{Belle:2023yoe}.  
Unlike in $\Lambda_b\to ph$ decays~\cite{Han:2025tvc, Han:2024kgz}, cancellations among different partial-wave CPVs do not occur here. Instead, the CPVs of the $S$- and $P$-wave components accumulate coherently, leading to the large total CPVs in $\Lambda_b \to \Lambda D_\pm$ decays. Neglecting $D^0-\bar{D}^0$ mixing and CP violation in $D$-meson decays~\cite{Grossman:2005rp}, the large CP-violating effects can be observed through the $\Lambda_b \to \Lambda (\to p\pi) D$ processes, with the $D$-meson reconstructed in either CP-even eigenstates ($K^+K^-,\pi^+\pi^-$) or CP-odd eigenstates ($K_S(\pi^0,\rho^0,\omega,\phi)$)~\cite{Giri:2001ju}. 
Given that the CPVs in modes with opposite CP parities differ by a sign, one can combine the modes with the same CP parity to enhance the statistics. 

{\it CP asymmetries  in the angular distributions---}
  CPV  can also be observed by studying the interference between the $S$- and $P$-wave contributions. Three decay parameters for the $\Lambda_b \to \Lambda D$ decay, which characterize the strengths of the $S$- and $P$-waves, are defined in terms of the partial-wave amplitudes as~\cite{Lee:1957qs}: 
\begin{align}
\alpha^\prime=-\frac{2\text{Re}(S^*P)}{|S|^2+|P|^2}, 
\beta^\prime=-\frac{2 \text{Im}(S^*P)}{|S|^2+|P|^2}, 
\gamma^\prime=\frac{|S|^2-|P|^2}{|S|^2+|P|^2},
\end{align}
which satisfy the identity, $\alpha^{\prime 2}+\beta^{\prime 2}+\gamma^{\prime 2}=1.$ 
Their relative ratios provide information about the relative strong phase between the $S$- and $P$-wave amplitudes. For example:
\begin{eqnarray}\label{eq:phase}
\sin(\delta_P-\delta_S)&=&-\frac{\beta^\prime}{\sqrt{1-\gamma^{\prime2}}},\nonumber\\
\cos(\delta_P-\delta_S)&=&-\frac{\alpha^\prime}{\sqrt{1-\gamma^{\prime2}}}. 
\end{eqnarray}

Measuring $\alpha^\prime$ does not require the initial  $\Lambda_b$ polarization. 
However, both the polarizations of the initial and final baryons are needed to extract $\beta^\prime$ and $\gamma^\prime$~\cite{Dunietz:1992ti}. Since $\Lambda_b$ baryons produced in $pp$ collisions at the LHC are unpolarized~\cite{LHCb:2020iux}, these decay parameters could be measured in the future at electron-ion colliders~\cite{Accardi:2012qut} and Z-factories (e.g. FCC-ee)~\cite{FCC:2018bvk}, where abundant polarized $b$-baryons are expected to be produced. 

The CP asymmetries associated with the decay parameters are defined as follows:
\begin{eqnarray}
A^{\alpha^\prime}_{CP}  = \frac{\alpha^\prime + \bar{\alpha^\prime}}{2}, 
A^{\beta^\prime}_{CP}   = \frac{\beta^\prime  +  \bar{\beta^\prime}}{2},
A^{\gamma^\prime}_{CP} = \frac{\gamma^\prime - \bar{\gamma^\prime}}{2},
\end{eqnarray}
The numerical results for these CP asymmetries are presented in Table.~\ref{tab:alpha}.  
It is observed that the CP asymmetries $A^{\beta^\prime}_{CP}$ are prominent for both the CP-even and CP-odd modes. Again, as there is no relative weak phase, the CP-violating effects associated with the decay parameters for both the $D^0$ and $\bar{D}^0$ modes are absent.

\begin{table}[!htbh]
	\caption{ The predicted  decay parameters and  associated $CP$ asymmetries of $\Lambda_b\rightarrow \Lambda D$ decays.}
	\label{tab:alpha}
	\begin{tabular}[t]{lcccc}
	\hline\hline
       Observable             & $\Lambda_b\rightarrow \Lambda D^0$  &$\Lambda_b\rightarrow \Lambda\bar{D}^0$ &$\Lambda_b\rightarrow \Lambda D_{+}$ &$\Lambda_b\rightarrow \Lambda D_{-}$  \\ \hline
   $\alpha^\prime$            &$-0.90^{+0.06}_{-0.06}$
                       &$-0.99^{+0.00}_{-0.01}$
                       &$-0.90^{+0.05}_{-0.05}$
                       &$-0.96^{+0.02}_{-0.03}$  \\
   $\beta^\prime$             &$-0.12^{+0.07}_{-0.10}$
                       &$0.05^{+0.06}_{-0.03}$
                       &$0.15^{+0.08}_{-0.11}$
                       &$-0.16^{+0.07}_{-0.08}$   \\
   $\gamma^\prime$     &$0.41^{+0.10}_{-0.15}$
                       &$0.11^{+0.01}_{-0.10}$
                       &$0.40^{+0.09}_{-0.11}$
                       &$0.22^{+0.06}_{-0.13}$     \\
   $A^{\alpha^\prime}_{CP}$     &$0$ &$0$      &$0.03^{+0.02}_{-0.01}$ &$-0.11^{+0.06}_{-0.06}$    \\
   $A^{\beta^\prime}_{CP}$      &$0$ &$0$      &$0.12^{+0.05}_{-0.05}$ &$-0.25^{+0.11}_{-0.09}$    \\
   $A^{\gamma^\prime}_{CP} $ &$0$     &$0$    &$0.08^{+0.05}_{-0.03}$ &$-0.18^{+0.06}_{-0.09}$    \\
		\hline\hline
	\end{tabular}
\end{table}

{\it Determination of $\gamma$ via $\Lambda_b\rightarrow \Lambda D$.---}
The CKM angle $\gamma$ is of particular interest because it can be measured in purely tree-level decays with negligible theoretical uncertainty~\cite{Brod:2013sga}. Gronau, London, and Wyler (GLW) proposed a method to extract $\gamma$ from $B \to D K$ decays using amplitude triangle relations~\cite{Gronau:1990ra, Gronau:1991dp}. This strategy is more directly applicable to the pure tree decays of $\Lambda_b \to \Lambda D$, where the interfering amplitudes are presumably of comparable size, as previously mentioned. In principle, one can constrain the angle $\gamma$ from the measurements of partial-wave CPVs via Eq.~\eqref{eq:CPV1}. However, a model-independent determination of $\gamma$ requires the measurement of these partial-wave observables and the associated strong-phase differences, which in turn necessitates complete information on $\alpha^\prime$, $\beta^\prime$, and $\gamma^\prime$ (see Eq.~\eqref{eq:phase}). As noted earlier, extracting $\beta^\prime$ and $\gamma^\prime$ remains challenging at present.


In this work, we propose a novel scheme to extract $\gamma$ in a model-independent way by combining angular distribution parameters and the decay rates in $\Lambda_b \rightarrow \Lambda D_{\pm}$ decays. We neglect  $D^0-\bar{D}^0$ mixing and CPV in $D$ decays due to their smallness~\cite{Grossman:2005rp, Wang:2012ie}. The parameter $\alpha^\prime_\pm$ for the CP-eigenstate modes can be written as
\begin{eqnarray}\label{eq:alpha}
\mathcal{A}_{\pm}\alpha^\prime_{\pm}=-2\text{Re}(S^*_{\pm}P_{\pm}),
\end{eqnarray}
where the amplitude squared $\mathcal{A}_{\pm}=|S_{\pm}|^2+|P_{\pm}|^2$ is proportional to the decay rate $\Gamma_\pm$ via Eq.~\eqref{eq:width}. Summing $\mathcal{A}_{\pm}\alpha_{\pm}$  with its counterpart in the CP-conjugate process yields
\begin{eqnarray}\label{eq:alpha1}
\mathcal{A}_{\pm}\alpha^\prime_{\pm}+\mathcal{\bar{A}}_{\pm}\bar{\alpha}^\prime_{\pm}=-2\text{Re}(S^*_{\pm}P_{\pm}+\bar{S}^*_{\pm}\bar{P}_{\pm}).
\end{eqnarray}
Substituting the partial-wave amplitudes from Eqs.~\eqref{eq:SP1}-~\eqref{eq:SPcpb} into Eq.~\eqref{eq:alpha1}, we obtain
\begin{eqnarray}\label{eq:ggg}
\sin\gamma=\frac{\pm(\mathcal{A}_{\pm}\alpha^\prime_{\pm}+\mathcal{\bar{A}}_{\pm}\bar{\alpha}^\prime_{\pm})}{2 (|S_D||P_{\bar D}|\sin\delta_{S_DP_{\bar D}}+|P_D||S_{\bar D}|\sin\delta_{P_DS_{\bar D}})}, 
\end{eqnarray}
with $\delta_{ij}=\delta_i-\delta_j$. To resolve the two-fold ambiguity in $\sin \gamma$ in Eq.~\eqref{eq:ggg}, we perform a subtraction in Eq.~\eqref{eq:alpha1}, yielding
\begin{eqnarray}\label{eq:gggp}
\cos \gamma=\frac{\pm(\alpha^\prime_D \mathcal{A}_D + \alpha^\prime_{\bar D} \mathcal{A}_{\bar D}-(\mathcal{A}_{\pm}\alpha^\prime_{\pm}-\mathcal{\bar{A}}_{\pm}\bar{\alpha}^\prime_{\pm}))}{2 (|S_D||P_{\bar D}|\cos\delta_{S_DP_{\bar D}} +|P_D||S_{\bar D}|\cos\delta_{P_DS_{\bar D}})}.
\end{eqnarray}
The same analysis on $\beta^\prime_\pm$ and $\gamma^\prime_\pm$ gives two more $\gamma$-extraction formulas
\begin{eqnarray}\label{eq:ggg1}
\sin \gamma &=& \frac{\pm(\mathcal{A}_{\pm}\beta^\prime_{\pm}+\mathcal{\bar{A}}_{\pm}\bar{\beta}^\prime_{\pm})}{2 [|S_D||P_{\bar D}|\cos\delta_{S_DP_{\bar D}}-|P_D||S_{\bar D}|\cos\delta_{P_DS_{\bar D}}]},\nonumber\\
\cos \gamma&=&\frac{\pm(\beta^\prime_D \mathcal{A}_D + \beta^\prime_{\bar D} \mathcal{A}_{\bar D}-(\mathcal{A}_{\pm}\beta^\prime_{\pm}-\mathcal{\bar{A}}_{\pm}\bar{\beta}^\prime_{\pm}))}
{2 (|P_D||S_{\bar D}|\sin\delta_{P_DS_{\bar D}}-|S_D||P_{\bar D}|\sin\delta_{S_DP_{\bar D}})}, 
\end{eqnarray}
and
\begin{eqnarray}\label{eq:ggg2}
\sin \gamma&=&\frac{\pm(\mathcal{A}_{\pm}\gamma^\prime_{\pm}-\mathcal{\bar{A}}_{\pm}\bar{\gamma}^\prime_{\pm})}{2 [|P_D||P_{\bar D}|\sin\delta_{P_DP_{\bar D}}-|S_D||S_{\bar D}|\cos\delta_{S_DS_{\bar D}}]},\nonumber\\
\cos \gamma&=&\frac{\mp(\gamma^\prime_D \mathcal{A}_D + \gamma^\prime_{\bar D} \mathcal{A}_{\bar D}-(\mathcal{A}_{\pm}\gamma^\prime_{\pm}+\mathcal{\bar{A}}_{\pm}\bar{\gamma}^\prime_{\pm}))}
{2 (|S_D||S_{\bar D}|\cos\delta_{S_DS_{\bar D}}-|P_D||P_{\bar D}|\cos\delta_{P_DP_{\bar D}})}. 
\end{eqnarray}
It is worth noting that the extraction formulas based on $\alpha^\prime_\pm$ and $\beta^\prime_\pm$ are complementary to each other in their strong phase dependence. This suggests that the uncertainty arising from the strong phase can be reduced if both quantities are measured simultaneously in experiments. Despite the challenges in measuring $\beta^\prime_\pm$ and $\gamma^\prime_\pm$, which require the initial polarization of $b$-baryons, these formulas provide valuable insights and allow us to extract $\gamma$ with different dependencies on the strong phases.

If the strong phases are obtained from model calculations (as shown in Tables~\ref{tab:amplitude1} and~\ref{tab:amplitude2}), it is sufficient to determine $\gamma$ by measuring the decay rates $\Gamma_{\pm}$ and the angular distribution parameters $\alpha^\prime_{\pm}$ for the decays $\Lambda_b \to \Lambda D_{\pm}$ and their CP-conjugate counterparts, using Eq.~(\ref{eq:ggg}). Notably, measurements of $\alpha^\prime_{\pm}$ are currently more experimentally accessible than those of $\beta^\prime_\pm$ and $\gamma^\prime_\pm$. 

With the LHCb Upgrade~I accumulating up to \(50\ \text{fb}^{-1}\) of data in the HL-LHC era~\cite{LHCb:2023hlw},
a complete angular analysis of \(\Lambda_b \to \Lambda(\to p\pi^-) D\) is expected to be achievable at LHCb Run~4~\cite{Geng:2022osc}. A determination of $\gamma$ with a precision of 1 degree or better is foreseen. 
The proposed extraction strategy offers a new pathway for constraining the CKM angle $\gamma$ with baryonic decays, and it can be readily adapted to the $\Xi_b \rightarrow \Xi D_{\pm}$  decays.

{\it Comparison with $\Xi_b\rightarrow\Xi D$ decays---}
  As noted in Ref.~\cite{Geng:2022osc}, the dynamics of $\Xi_b\to\Xi D$ are simpler than those of $\Lambda_b\to\Lambda D$ due to the absence of the $C'$ and $E$ topologies. Consequently, the positive interference between $C$ and $C'$ in $\Xi_b\to\Xi \bar{D}^0$ is absent, and the corresponding amplitude ratios are strongly suppressed by the relevant CKM factors. For $\Xi_b\to\Xi D$ decays, we predict the amplitude ratios $r_S = 0.29$ and $r_P = 0.77$, along with small relative strong phases: $\delta_{S_D}-\delta_{S_{\bar D}}=0.10$ and $\delta_{P_D}-\delta_{P_{\bar D}}=0.12$. These two features imply that the total CPVs are less than $10\%$ for both CP-even and CP-odd modes (see Table~\ref{tab:obserxi}). We therefore encourage the LHCb collaboration to prioritize $\Lambda_b\to\Lambda D$ decays as golden channels for CPV searches and CKM metrology in the baryon sector.

\begin{table}[!htbh]
	\caption{ Same as Table~\ref{tab:obser} but for the $\Xi_b\rightarrow \Xi D$  decays. The CPVs are given in $\%$.}
	\label{tab:obserxi}
	\begin{tabular}[t]{lcccc}
	\hline\hline
   $D$  & $\mathcal{B}(10^{-5})$ & $A_{CP}$   & $\kappa A^S_{CP}$ &$(1-\kappa) A^P_{CP}$ \\ \hline
 $D_{+}$      &$2.8^{+1.8}_{-1.5}$  &$-5.3^{+2.0}_{-2.0}$   &$-3.2^{+3.0}_{-0.2}$   &$-2.1^{+0.1}_{-2.9}$ \\
 $D_{-}$      &$2.2^{+1.2}_{-1.1}$  &$7.9^{+3.1}_{-2.8}$
  &$4.7^{+0.4}_{-4.4}$     &$3.2^{+4.4}_{-0.1}$ \\
		\hline\hline
	\end{tabular}
\end{table}

{\it Conclusion---} 
Motivated by the recent LHCb observation of CPV in baryon decays, we present the first complete QCD analysis of color-suppressed $\Lambda_b \to \Lambda D$ decays in the perturbative QCD framework, including nonfactorizable contributions from internal $W$-emission and $W$-exchange. These effects generate nonzero strong-phase differences and enhance interfering amplitude ratios for both $S$- and $P$-wave components, leading to large partial-wave CPVs. We show that these partial-wave CPVs share the same sign and accumulate, yielding significant total CPVs. The predicted CPVs for $\Lambda_b \to \Lambda D_{\pm}$ can reach $50\%$, large enough for experimental detection. We also provide first high-precision predictions for CP-violating observables in angular distributions. Furthermore, we propose a novel strategy to extract $\gamma$ by combining angular distribution parameters and decay rates for $\Lambda_b \to \Lambda (D^0,\bar{D}^0,D_{\pm})$ and their charge conjugates; this strategy also applies to $\Xi_b \to \Xi D_{\pm}$ decays.

{\it Acknowledgments---} 
We would like to acknowledge Wenbin Qian, Fu-Sheng Yu,  Chia-Wei Liu, and Qin Qin for valuable discussions. Special thanks go to Hsiang-nan Li for reviewing an earlier draft of the manuscript and for his constructive comments.
This work is supported in part by the National Science Foundation of China under the Grants Nos. 12375089, 12435004, and 12075086, and the Natural Science Foundation of Shandong province under the Grant No. ZR2022ZD26 and ZR2022MA035.

\end{document}